\documentclass[useAMS,usenatbib,usegraphicx]{mn2e}

\def\mpc{\,{\rm Mpc}} 
\def\hmpc {\,h^{-1}\,{\rm Mpc}} 
\def\hunit {\, {\rm km \, s^{-1} \,Mpc^{-1} } } 
\def\zacc {z_{\rm acc}} 

\def\simlt{\mathrel{\lower0.6ex\hbox{$\buildrel {\textstyle <} 
  \over {\scriptstyle \sim}$}}}

\def\simgt{\mathrel{\lower0.6ex\hbox{$\buildrel {\textstyle >}
 \over {\scriptstyle \sim}$}}}

\def\newtwo { } 
\def\newthree{ } 

\title[A cosmic speed-trap]
 {A cosmic speed-trap: a gravity-independent test of cosmic
  acceleration using baryon acoustic oscillations}
\author[Will Sutherland] 
{Will Sutherland$^{1}$\thanks{E-mail: w.j.sutherland@qmul.ac.uk} 
\\
$^{1}$Astronomy Unit, Queen Mary University of London, 
  Mile End Road, London E1 4NS 
} 

\begin{document}

\date{Accepted by MNRAS, 2011 Nov 9. Received 2011 Nov 9; in original
  form 2011 May 3} 

\pagerange{\pageref{firstpage}--\pageref{lastpage}} \pubyear{2011}

\maketitle

\label{firstpage}

\begin{abstract}
We propose a new and highly model-independent test of 
 cosmic acceleration by comparing observations of 
 the baryon acoustic oscillation (BAO) scale at low and intermediate
 redshifts: 
 we derive a new inequality relating BAO observables at
 two distinct redshifts, which must be 
 satisfied for any reasonable homogeneous non-accelerating model, 
 but is violated by models similar to $\Lambda$CDM, due to 
  acceleration in the recent past. 
  This test is fully independent of
 the theory of gravity (GR or otherwise), the Friedmann equations, 
 CMB and supernova observations: the test  
  assumes only the Cosmological Principle, and that the
  length-scale of the BAO feature is fixed in comoving coordinates.  
 Given realistic medium-term observations from BOSS, this test is
 expected to exclude all homogeneous 
 non-accelerating models at $\sim 4\sigma$ significance,
 and can reach $\simgt 7\sigma$ with next-generation surveys.  
\end{abstract}


\section{Introduction}
\label{sec:intro} 

In the last 10--15 years, the $\Lambda$CDM model has
 been established as the standard model of large-scale cosmology;
 the model is an excellent match to many observations
 including the anisotropies in the CMB measured by WMAP 
 \citep{komatsu11} and
 other experiments, 
 the large-scale clustering of galaxies \citep{perc10}, 
 the Hubble diagram for high-z supernovae \citep{guy10,conley11},
  and the abundance
 and baryon fraction of rich clusters of galaxies \citep{allen11}. 

Despite these great observational successes, the model appears unnatural
since 96\% of the universe's mass-energy is not observed, but 
is only inferred from fitting the observations. 
 Also, the dark sector contains at least two apparently unrelated 
 components, dark matter and dark energy; 
 recent reviews of dark energy are given by \citet{fth08} and
 \citet{linder08}. 

The most direct evidence for cosmic acceleration comes
 from the Hubble diagram of Type-Ia supernovae \citep{guy10, conley11}, 
 which shows that SNe at $0.3 \simlt z \simlt 0.9$ are fainter, 
 relative to local SNe, than can be accommodated in any 
 Friedmann-Robertson-Walker model without dark energy. 
 A model-independent approach has also been given by \citet{st06}, 
 who show that the SNe results require accelerated expansion
 at $z < 0.4$ at around the 
 $5\sigma$ significance level 
 {\em without} assuming the Friedmann equations. 

However, there are some possible loopholes in the supernova
 results: since they are fundamentally based on brightness measurements,
 the interpretation could be affected by 
 either unexpected evolution of the mean SNe properties over cosmic time,
 or some process which removes photons en route to our
 telescopes, such as peculiar dust 
 or more exotic effects such as photon-dark matter interactions. 
 The simplest such effects with monotonic time-dependence 
  are strongly disfavoured by SN observations at $z > 1$ \citep{riess07}, 
  but more complex time-dependent effects 
 could still leave these loopholes open. 

 Independent of supernovae, there is powerful support for 
 dark energy from observations of the anisotropies in the cosmic 
 microwave background \citep{larson10, komatsu11} and 
 the large-scale clustering of galaxies \citep{perc10}, but this 
  is dependent on assuming general relativity and
 the Friedmann equations; if these both hold, the
 model parameters are tightly constrained by CMB and LSS data, 
  and the expansion history $a(t)$ must match $\Lambda$CDM 
 models within a few percent.  
 However, in alternative
  gravity theories, we cannot make model-independent statements 
  from the CMB or large-scale structure:   clearly
 any successful modified-gravity model should eventually
 be consistent with these observations,  
 but the model space of modified gravity is large and the 
 calculations non-trivial; so in non-GR models we cannot necessarily 
  use the CMB and LSS observations to make any definite statement  
 about recent acceleration.  

The accelerated expansion is so startling  
 that it is desirable to test it via multiple routes 
 with a minimum number of model assumptions. 
A very direct test of acceleration has been proposed using 
 the ``cosmic drift'', which is the small change in redshift 
 for fixed object(s) over time (e.g. \citealt{liske08}); 
 the predicted change is $dz/dt = (1+z) H_0 - H(z)$.  
 However, this effect is tiny over human timescales, of order cm/s/year, 
 and will probably require over 20 years baseline to get a 
 convincing detection. 

 Here we propose a new and robust test for cosmic acceleration based
  only on the cosmic ``standard ruler'' in the galaxy correlation
 function: in the standard model, this is a feature 
 created by acoustic oscillations in the baryon-photon fluid
 before recombination (e.g. \citealt{peeb-yu70}); this was analysed in 
  more detail by \citet{eh98} and \citet{mwp99}, then 
  first detected in 2005 by \citet{eis05} in SDSS data, 
  and \citet{cole05} using the 2dFGRS survey.  
 The length of this ruler, hereafter $r_s$, depends
 only on matter and radiation densities and is accurately
 predicted from CMB observations at $\approx 153 \pm 2 \mpc$
 \citep{komatsu11}.  
 Many recent studies (e.g. \citealt{esw07}, \citealt{shoji09}, 
 \citealt{abdalla10}, \citealt{tian10}) 
  have shown how precision measurements
 of this BAO scale from huge galaxy redshift surveys can provide
 powerful constraints on the properties of dark energy, and
 test for evolution of dark energy density; more details 
 are given in Section~\ref{sec:bao}. 

However, in the current paper we do {\bf not} assume any gravity theory 
  or the actual length scale of this feature, only that we can 
 observe some feature at a specific lengthscale imprinted on the galaxy 
 distribution at  high redshift, which expands with the Hubble
 expansion and remains a constant ruler in comoving
 coordinates.  
 We then derive an inequality relating observations comparing
 this ruler at low
 and intermediate redshift, which is satisfied in {\em  any reasonable} 
 non-accelerating model, but is violated by accelerating models 
 approximating $\Lambda$CDM. 
 In more detail, we use the {\em radial} component
 of the BAO feature at $z_2 \sim 0.75$ to constrain the product 
 $H(z_2) \,r_s$, 
 and we then compare to the {\em spherical-averaged} BAO 
  feature at low redshift  $z_1 \sim 0.2$, 
 which is related to the average of $1/H(z)$ 
  at $0 \le z \le z_1$.  Then, assuming any 
  non-accelerating model we derive 
 a strict upper limit on the ratio of these.  
 Models approximating standard $\Lambda$CDM predict a result which violates 
  this inequality by a substantial amount 
 $\sim 10 - 20 \%$, depending on cosmological parameters and redshift. 
 Future large redshift surveys should be able to measure this ratio 
 to $\le 2\%$ precision: 
 assuming our inequality is significantly violated as predicted, 
 we can then exclude all homogeneous non-accelerating models 
 regardless of Friedmann equations, gravity theory or details of the 
  expansion history.  

The plan of the paper is as follows: in \S~\ref{sec:bao} 
 we review the basic features and observables of baryon acoustic
 oscillations. In \S~\ref{sec:trap} we derive the new inequality
 relating BAO observables for non-accelerating models.
 In \S~\ref{sec:disc} we discuss future observations
 and related issues,  and we summarise
 our conclusions in \S~\ref{sec:conc}.

\section[]{Observations of the BAO feature}
\label{sec:bao} 

 The baryon acoustic oscillation (hereafter BAO) feature \citep{eh98, mwp99} 
  is a bump in the galaxy correlation function $\xi(r)$,
 or equivalently a decaying series of wiggles in the power spectrum $P(k)$, 
 corresponding to a comoving length denoted by $r_s$, created
 by acoustic waves in the early universe prior to decoupling. 
 (See \citet{bh10} for a recent review).   
 In the standard model, its length-scale is 
 essentially set by the distance that a sound wave can propagate 
 prior to the ``drag epoch''
 at $z_d \approx 1020$, denoted $r_s(z_d)$, and this length depends only
 on physical densities of matter $\Omega_m h^2$ and
  baryons $\Omega_b h^2$, 
 (together with radiation density $\Omega_r h^2$ which is pinned 
   very precisely by the CMB temperature). 
 In the standard model the relative heights of the acoustic peaks
  in CMB anisotropies constrain $\Omega_m h^2$ and 
 $\Omega_b h^2$ well \citep{komatsu11},
 which leads to a prediction  $r_s \approx 153 \mpc$ comoving
 with approximately 1.5 percent precision. This predicted length 
 does not rely on the assumption of a flat universe, 
  since the relative CMB peak heights constrain 
 the various densities reasonably well without assuming flatness. 
 However, the CMB-predicted length $r_S$ does depend on assuming standard GR, 
  and several assumptions about
 the mass-energy budget including standard neutrino content, negligible
 early dark energy, no late-decaying dark matter, 
  {\newtwo negligible admixture of isocurvature perturbations}, etc.   
 However, in the rest of this paper we leave $r_s$ 
 as an arbitrary comoving scale, which cancels later.  

 The BAO feature provides a standard ruler which 
 can be observed at low to moderate
  redshift using very large galaxy redshift surveys; 
  in the small angle approximation and
 assuming we observe a redshift shell which is thin compared
 with its mean redshift $z$, there are two primary observables
  derived from a BAO survey: firstly  
 the angle on the sky subtended by the BAO feature transverse
  to the line of sight,  
  $\Delta\theta(z) = r_s / [(1+z) D_A(z)]$, 
 where $D_A(z)$ is the conventional (proper) angular-diameter distance
  to redshift $z$; 
 and secondly the difference in redshift along
  one BAO length along the line of sight is 
 $\Delta z_{/\!/}(z) = r_s H(z) / c $ (e.g. \citealt{bg03}, 
 \citealt{seo03}). 
 We note that calculating comoving galaxy separations from
  observed positions and redshifts requires a reference cosmology, hence
 a difference between the true and reference cosmology will produce
 an error in the inferred $r_s$; 
 however, any error in the reference model cancels to first order
  in the dimensionless ratios $r_s/D_A(z)$ and $r_s \,H(z) /c $, 
 so both of these ratios can be well constrained 
 with minimal theory-dependence by measuring BAOs in a 
 galaxy redshift survey.  
 
 The ability to independently probe 
  $D_A(z)$ and $H(z)$ is a powerful advantage of BAOs over
 other low-redshift cosmological tests. Furthermore, a redshift
 survey useful for BAOs can also measure growth of structure 
 via redshift-space distortions and thus test for consistency with GR,
 though we do not consider this here. 

However, in practice, current galaxy redshift surveys are not quite
 large enough to robustly measure the BAO feature separately
 in angular and radial directions (though there are tentative
 detections, e.g. \citealt{gazta09} ).  The current measurements
 primarily constrain a spherically-averaged scale, 
 called $D_V$,  which is defined by \citet{eis05} as 
\begin{equation}
 \label{eq:dv} 
  D_V(z) \equiv \left[ (1+z)^2 D_A^2(z) \frac{c z}{H(z)} \right]^{1/3} \ ; 
\end{equation} 
 this is essentially a geometric mean of two transverse directions and
  one radial direction. 
 Observations using the 2dFGRS and SDSS-II redshift surveys
 have measured the dimensionless ratio 
  $d(z) \equiv r_s/D_V(z)$ at low redshifts 
 \citep{perc10,kazin10}, which we discuss later.  
 We note that as $z \rightarrow 0$, $D_V(z) \rightarrow c z / H_0$; 
   however, this approximation is not very useful 
 in practice, since we cannot measure
 the BAO feature at very low redshift $z < 0.02$ where corrections of
 order $z^2$ are unimportant.   We give a better approximation below 
 in \S~\ref{sec:dvapprox}.   

In practice, the BAO feature is not a sharp spike but a hump in 
 $\xi(r)$ of width approximately 15\% of $r_s$, 
 so there are several subtle effects in actually
 extracting the scale $r_s$ from a redshift survey:  
 we discuss these in more detail in \S~\ref{sec:rshift}. 
 However, 
 for the purposes of this paper we only
 need to assume that $r_s$ is a constant comoving length to 
 $\sim 1\%$ at redshift $\le 0.8$, so these precision details are 
 relatively unimportant for the rest of this paper. 

\section{The cosmic speed trap} 
\label{sec:trap} 

Here we derive a new inequality which we denote the ``cosmic
 speed-trap'', which must be satisfied by any reasonable
 non-accelerating model, but is violated by $\Lambda$CDM
 and other accelerating models.   We start off by assuming
 an arbitrary non-accelerating model,  and deriving a lower
 limit for $D_V(z_1)$ in terms of the value of $H(z_2)$
 at a {\em higher} redshift $z_2$.  Then, we form a ratio of
 BAO observables which eliminates $H(z_2)$ and $r_s$, and we
 obtain the speed-trap inequality (\ref{eq:xs})
  which forms our main new result.  

\subsection{An inequality for $D_V$ in non-accelerating models} 
\label{sec:dvlim} 
Here we derive an inequality for $D_V(z)$ which is satisfied
 in any non-accelerating model, but may be violated by 
 acceleration. 

First we define as usual $a$ to be the cosmic expansion factor
 relative to the present day with $a_0 = 1$, redshift 
 $z$ by $1 + z \equiv a^{-1}$, 
 and the Hubble parameter $H(a) = \dot{a}/a$ where dot represents
 time derivative. 
 Then we have the expansion rate 
\begin{equation} 
 \dot{a} = a H(a) = \frac{H(z)}{1+z} \ ; 
\end{equation}
  if the expansion of the universe was non-accelerating,
  then $\ddot{a}$ is non-positive and the function above must
 be non-increasing with time or $a$, therefore non-decreasing with
 increasing $z$.  Therefore, if we consider
 any two redshifts $z_1 < z_2$, in any non-accelerating universe,  
\begin{equation} 
\label{eq:hineq} 
  \frac{H(z_1)}{1+z_1}  \le \frac{H(z_2)}{1+z_2} \ .  
\end{equation} 
Assuming only the cosmological principle, 
any observed violation of this inequality is a direct proof
 that the expansion has accelerated, on average, 
  between the earlier epoch $z_2$ and the later epoch $z_1$,
 without reference to any specific theory of gravity or geometry.   

\begin{figure} 
\includegraphics[width=6cm, angle=-90]{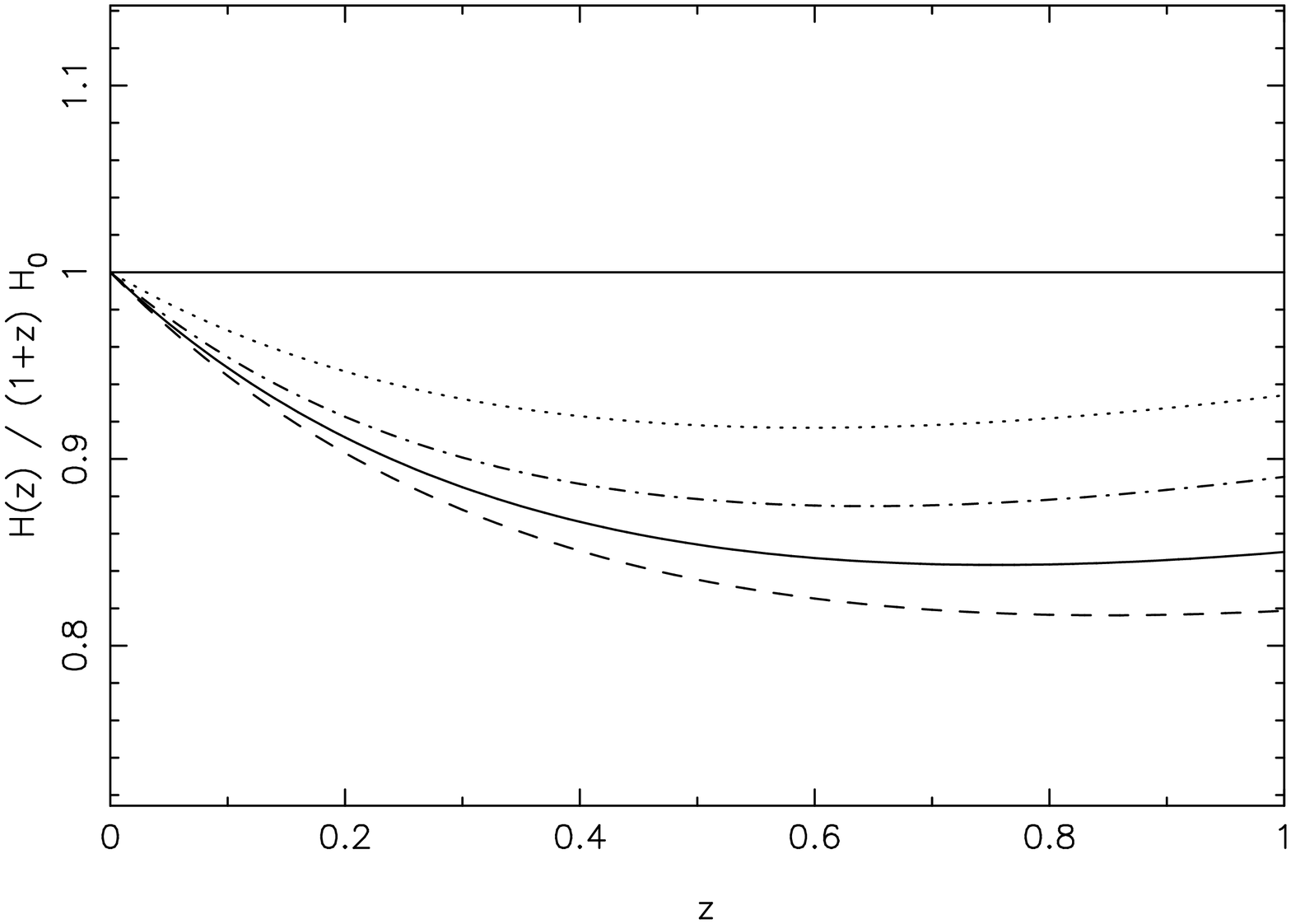}  
\includegraphics[width=6cm, angle=-90]{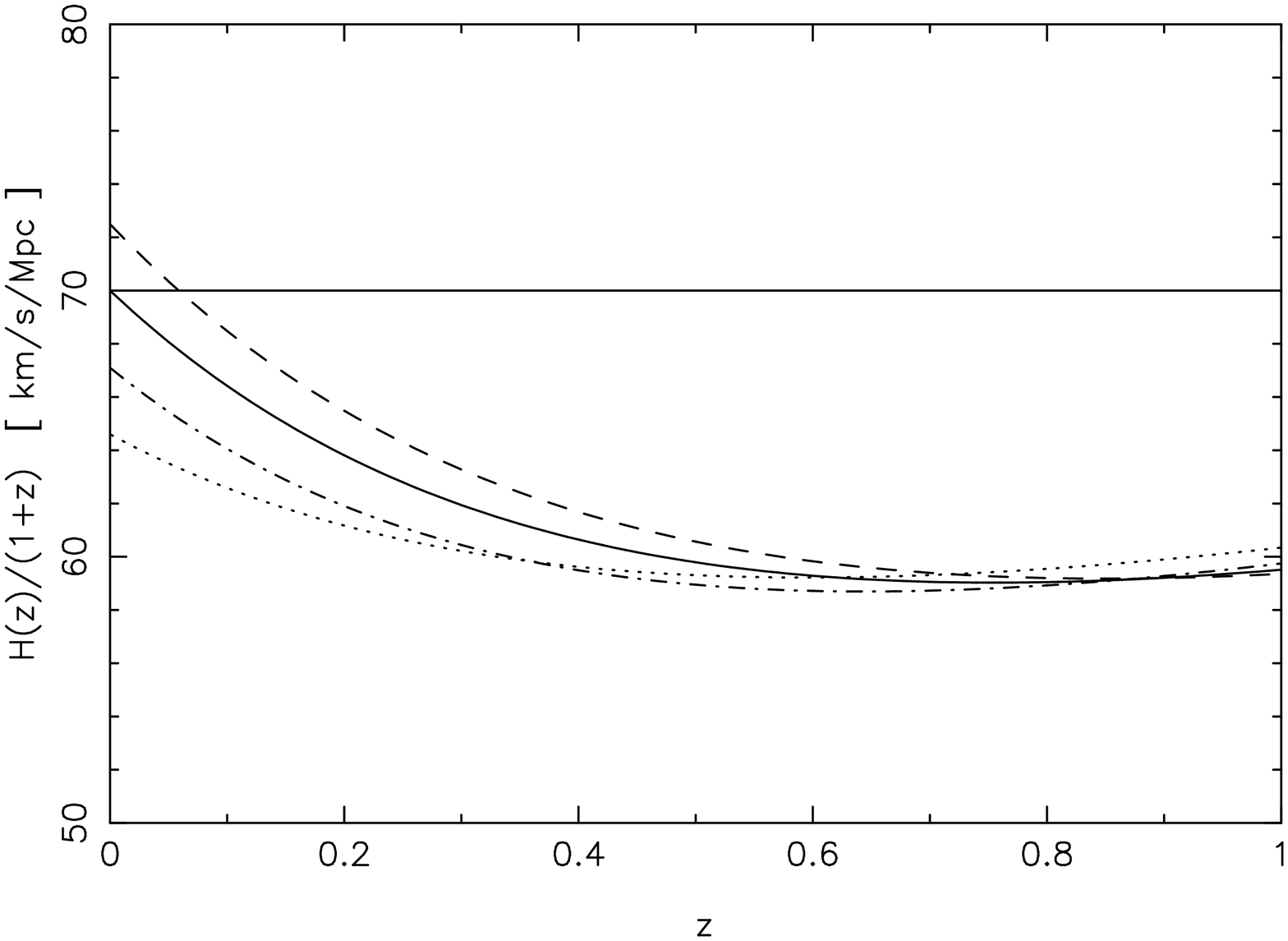}  
\caption{The expansion rate $\dot{a} = H(z)/(1+z)$ from the
 standard Friedmann equation is shown 
 as a function of redshift, for four models: three flat $\Lambda$CDM with
 respectively $\Omega_m = 0.24$ (dashed), 0.27 (solid) and 0.31 
 (dot-dash); also one wCDM model with 
 $\Omega_m = 0.32$, $w = -0.85$ (dotted).  
 The upper panel shows expansion rate relative to 
  present-day $H_0$, while the lower
 panel shows absolute expansion rate, where for each model $H_0$ 
  is adjusted to keep the CMB acoustic angle constant. 
 Parameters are given in Table~\ref{tab:models}. 
  }  
\label{fig:hz} 
\end{figure}

A concordance $\Lambda$CDM model {\em does} 
 violate this inequality due to the recent positive acceleration: 
 a minimum value of 
 $H(z)/(1+z)$ occurred at $\zacc = 
  \sqrt[3]{2\,\Omega_\Lambda /\Omega_m} - 1$; for
  the concordance value $\Omega_m \approx 0.27$, this gives 
  $\zacc \approx 0.75$, 
 and $H(\zacc) / (1 + \zacc) \approx 0.85 \, H_0$. 
 The expansion rate $H(z)/(1+z)$ is shown in Figure~\ref{fig:hz} for a few
  representative models: 
 it is notable that the value of $H(z)/(1+z)$ remains within a few
 percent of its minimum between $0.5 \le z \le 1.2$, 
  and it rises rather sharply at low redshift; for the concordance
 model it only crosses the half-way value between the 
 minimum and the present-day $H_0$ at the 
 modest redshift of $z \approx 0.17$, and three-quarters of
 the speedup has occurred since $z \approx 0.31$. 
 Thus the actual speedup of the expansion rate is
  quite concentrated at rather low redshift; this 
  becomes relevant later. 
 
\begin{table} 
\begin{tabular}{|l|c|c|c|c|} 
\hline 
Model  & $\Omega_m$ & $H_0$ &  $w$  &  $t_0$ \\ 
       &            & ($\hunit$) &  & (Gyr)  \\ 
\hline
C  & 0.27 & 70.0  &  $-1$  &  13.86  \\ 
L  & 0.24 & 72.5 &   $-1$  &  13.82  \\ 
H  & 0.31 & 67.1 &   $-1$  &  13.91  \\ 
W  & 0.32 & 64.6 &  $-0.85$ & 13.98 \\ 
\hline
\end{tabular}
\caption{Cosmological parameters for the four example 
  models discussed in the text; model C is the
  baseline concordance model, while the others are selected
  to roughly span the current $2\sigma$ allowed 
  range in $\Omega_m$ and $w$. 
  All are flat, and have $H_0$ adjusted to give 
  very similar values of $\ell_A$ consistent with WMAP,  
   therefore have similar values of $t_0$. } 
\label{tab:models}
\end{table}

Next, we suppose we have a measurement of $H(z_2)$ at
 an earlier epoch $z_2$; for a non-accelerating model 
 we now derive
 a {\em lower limit} on $D_V(z_1)$ at a later 
 epoch $z_1$ where $z_1 < z_2$.  

The comoving radial distance to redshift $z_1$ is 
\begin{equation} 
\label{eq:dr} 
  D_R(z_1) = c \int_0^{z_1} \frac{1}{H(z)} \, dz \quad . 
 \end{equation} 
If the universe is non-accelerating and $z_1 < z_2$, we
 can rearrange inequality (\ref{eq:hineq}) into
  $1/ H(z) \ge (1+z_2) / [H(z_2) (1+z)] $; inserting this
 we have 
\begin{equation} 
  D_R(z_1) \ge { c \,(1+z_2) \over H(z_2)} \ln(1+z_1)  
\end{equation} 

The {\newtwo proper} angular-diameter distance $D_A(z)$ is defined by 
\begin{equation} 
\label{eq:da} 
 (1+z) D_A(z) \equiv \vert R_C \vert \, 
   S_k \left( {D_R(z) \over \vert R_C \vert } \right) 
   = D_R(z) {S_k(x) \over x} 
\end{equation} 
 where $\vert R_C \vert $ is the curvature radius of the 
 universe in comoving Mpc, $x \equiv D_R(z)/ \vert R_C \vert$, and the 
 function $S_k(x)$ = $\sinh x, \ x , \ \sin x$ for 
 the cases $k = -1, \, 0, \, +1$ where $k$ is the sign of the curvature.  

Note that in the above we have left $R_C$ as a constant but arbitrary 
 curvature radius, thus we have {\em not} assumed the 
 Friedmann equation which gives 
 $R_C = (c/ H_0) \sqrt{ k / (\Omega_{tot} - 1) } $; 
  we have only assumed that the universe has a metric with 
 some well-defined curvature radius $R_C$,
  which follows from the assumption of homogeneity and 
 isotropy \citep{peacock}. 
 Also, we have not assumed any functional form for $H(z)$, only
 that it obeys the non-acceleration condition (\ref{eq:hineq})
 at all $z \le z_2$;  what happened earlier at $z \ge z_2$ is 
 immaterial. 

For the other term in $D_V$, we use a similar inequality 
 for $1/H(z_1)$ as above, which is
 \begin{equation} 
 {c z_1 \over H(z_1)} \ge { c z_1 (1+z_2) \over H(z_2) (1+z_1)} \ ; 
 \end{equation}  
 substituting both of the above into Eq.~\ref{eq:dv}, 
 we obtain the inequality 
\begin{equation} 
\label{eq:dvineq} 
 D_V(z_1) \ge {c (1+z_2) \over H(z_2) } \left[ z_1 \,(\ln(1+z_1))^2 \over 
  1 + z_1 \right]^{1/3} \, \left({S_k(x_1) \over x_1}\right)^{2/3} \; .  
\end{equation} 
 where $x_1 \equiv D_R(z_1) / \vert R_C \vert$ as above. 

This inequality is strict for any non-accelerating and homogeneous universe 
 with a Robertson-Walker
 metric, independent of details of the expansion history or
 the gravity model. This is not so useful on its own, but we will
 see in the next section how to combine observables to 
 cancel the $z_2$ dependence.  

We note that the factor $(S_k(x_1)/x_1)^{2/3} = 1$ exactly
 for flat models, and is $\ge 1$ for open models (so open models
 always strengthen the inequality); 
 the factor is $\le 1$ for closed models which weakens our 
 inequality, but only by a small amount if we consider 
 sufficiently low redshift $z_1$,  
 since the effect of curvature on distances only enters to third
 order in $z$;  
   at small $x$ and $k = +1$ we have 
\begin{equation}  
  \left( {S_k(x) \over x}\right)^{2/3} \approx 1 - \frac {x^2}{9} 
 \nonumber 
\end{equation} 
 therefore we need an upper limit on $x$ for closed models. 
{\newtwo 
 We get a firm limit as follows, using an upper bound on $D_R$ and a lower
  bound on $R_C$ for closed models.  

 To limit $D_R$, we can use the non-acceleration inequality (\ref{eq:hineq})  
 between $z = 0$ and an upper redshift $z_1$ 
  to get $1/H(z) \le 1/[H_0(1+z)]$, which now leads to 
 an {\em upper} bound on $D_R(z_1)$ in terms of $H_0$, 
  $D_R(z_1) \le (c/H_0) \ln(1+z_1)  \le c z_1 / H_0 $, for any non-accelerating
  model.  This gives $x_1 \le c z_1 / H_0 R_C$. 

 We may also obtain a lower bound on $R_C$ as follows:  
  in a closed model, it is clear
  from Eq.~\ref{eq:da} and $\sin x \le 1$ 
  that $D_A(z)$ cannot exceed $R_C / (1+z)$ 
  regardless of the expansion history $H(z)$.
  If we take for example $R_C = 0.6 \, c /H_0$ and 
   $H_0 = 70 \hunit$,  
  this leads to $D_A(z = 3) \le 642 \, {\rm Mpc}$,
   only $0.4\times$ the concordance value of $1638 \,{\rm Mpc}$.  
  However, observed angular sizes of $z \sim 3$ galaxies already convert to 
  rather small physical
 sizes based on the concordance model, and making them smaller by another
  factor $< 0.4$ appears to be seriously discrepant. We therefore exclude
  closed models with $R_C < 0.6 \,c/H_0$. 

 A stronger lower bound may be obtained with other methods:
 e.g. the luminosity distance $D_L(z = 1.5)$ measured from SNe 
 \citep{riess07} agrees
  well with the concordance model, and if we adopt a lower bound 
  $0.8 \times$ the concordance value, we obtain $R_C \ge 0.84 \, c/H_0$.  
 However, to remain fully independent
 of SNe data we do not use this below.  
 A stronger limit should also be possible 
 in future using angular BAO measurements
  at $z \sim 3$, e.g. from the HETDEX or BOSS projects.  

 However, for the following we take
   $R_C \ge 0.6 \, c/H_0$ as a conservative 
   gravity-independent lower limit for closed models. 
 This leads to a firm upper limit $x \le \ln(1+z) / 0.6$
  for closed non-accelerating models, which we use below.  
 } 

\subsection{The observable speed-trap}
\label{sec:xs} 

The above inequality (\ref{eq:dvineq})  
 relates the volume-distance $D_V(z_1)$ at low redshift to 
 the Hubble constant $H(z_2)$ at a higher redshift.  
 Neither of these
 quantities are directly observable at present, but it is possible to
  measure both of them relative to the BAO 
 length-scale $r_s$; then, dividing  
  these two cancels the length scale $r_s$ and 
 gives a ratio measurement. Applying 
 the $D_V$ inequality above gives us a limit which must be satisfied
 by any reasonable non-accelerating model,
 but is found to be violated by an expansion history close
 to $\Lambda$CDM, for a range of suitable choices 
 of $z_1 \sim 0.2, \; z_2 \sim 0.75$.  

The Hubble parameter $H(z_2)$ may be measured using
 the {\em radial} BAO scale (along the line of sight) 
 in a redshift shell near $z_2$; 
 for a thin shell and ignoring redshift-space distortion 
 effects, this gives the observable 
\begin{equation}
  \Delta z_{/\!/}(z_2) = \frac{r_s H(z_2)}{c} 
\end{equation} 
 In practice it is useful to divide by $1+z_2$ and define
\begin{equation} 
  y(z_2) \equiv { \Delta z_{/\!/}(z_2) \over 1 + z_2 } 
  = {  r_s \, H(z_2) \over c \,(1 + z_2) }  
\end{equation}
  since this $y$ is rather close to a constant over a substantial
 range of redshift in a $\Lambda$CDM model (as in Figure~\ref{fig:hz}), 
 and we will see that it has a convenient cancellation below.  

Using the SDSS-II redshift survey, \citet{perc10} have
 already measured the dimensionless ratio 
 \begin{equation} 
 d(z) \equiv r_s / D_V(z)
\end{equation}
  at redshift $z = 0.2$ and 0.35, and also a combined
  ratio at $z = 0.275$. (We discuss the numerical results later). 

\vspace{5mm} 
 We now form the ratio of observables $z_1 d(z_1) / y(z_2)$ 
 which gives,  from the definitions above 
\begin{equation} 
\label{eq:dovery} 
 { z_1 d(z_1) \over y(z_2) }  =  { c (1+z_2) \over H(z_2)} 
  {z_1 \over D_V(z_1) }  \qquad ; 
\end{equation} 
assuming only that $r_s$ is a fixed comoving ruler independent of $z$.  

 If we now assume that the universe has never accelerated 
 below redshift $z_2$,    
 we may apply the inequality (\ref{eq:dvineq}) for $D_V(z_1)$;
  this cancels the $z_2$ factors, giving the inequality  
\begin{equation} 
\label{eq:ydlim}
{ z_1 \, d(z_1) \over y(z_2) } \le   
  \left[ {z_1^2 (1 + z_1) \over (\ln(1+z_1))^2 }  \right]^{1/3} 
  \left({x_1 \over S_k(x_1)}\right)^{2/3}  \quad . 
\end{equation}  

It is more convenient to rearrange this to put the square-bracket term
 on the LHS, and define the quantity $X_S$ (``excess speed'') by 
\begin{equation}
\label{eq:xs} 
  X_S(z_1,z_2)  \equiv { z_1 \, d(z_1) \over y(z_2) }   
  \left[ (\ln(1+z_1))^2 \over {z_1^2 (1 + z_1) } \right]^{1/3} 
 \le   \left({x_1 \over S_k(x_1)}\right)^{2/3}  \, ;  
\end{equation} 
 where $X_S$ is a ratio of observables, and $x_1 = D_R(z_1)/R_C$
 as before.   
 (Note one may cancel some powers of $z_1$ on the LHS, but leaving them as 
 above makes both terms in $X_S$ well-behaved 
 as $z_1 \rightarrow 0$ .) 

This inequality forms the main result of our paper, our
  {\bf cosmic speed-trap}, 
 which must be obeyed for any chosen values $z_1$ and 
  $z_2$ with $z_1 \le z_2$,  
  given the following conditions: 
 \begin{enumerate} 
 \item The universe is nearly homogeneous  and isotropic
    with a Robertson-Walker metric. 
 \item The redshift is due to cosmological expansion and $c$ is constant.  
 \item   $r_s$ is the same comoving length at $z_1$ and $z_2$, and 
 \item The expansion has never accelerated in the
  interval $0 < z < z_2$. 
\end{enumerate} 
{\newtwo 
 If the speed-trap is observationally violated, 
  $X_S > (x_1/S_k(x_1))^{2/3}$
 at high significance, one or more
 of assumptions (i)-(iv) above must be false, independent of
 gravity theory or Friedmann equations. 
 To apply this test, 
  we also require an upper bound on the RHS, i.e. 
  an upper bound on $x_1$ for closed models,
 which we derive below 
 {\newthree (this is not strictly a 
  fifth ``assumption'', since it 
 follows from observational data assuming (i), (ii) and (iv) above). 
 }
}     

\begin{figure*}  
\includegraphics[width=12cm, angle=-90]{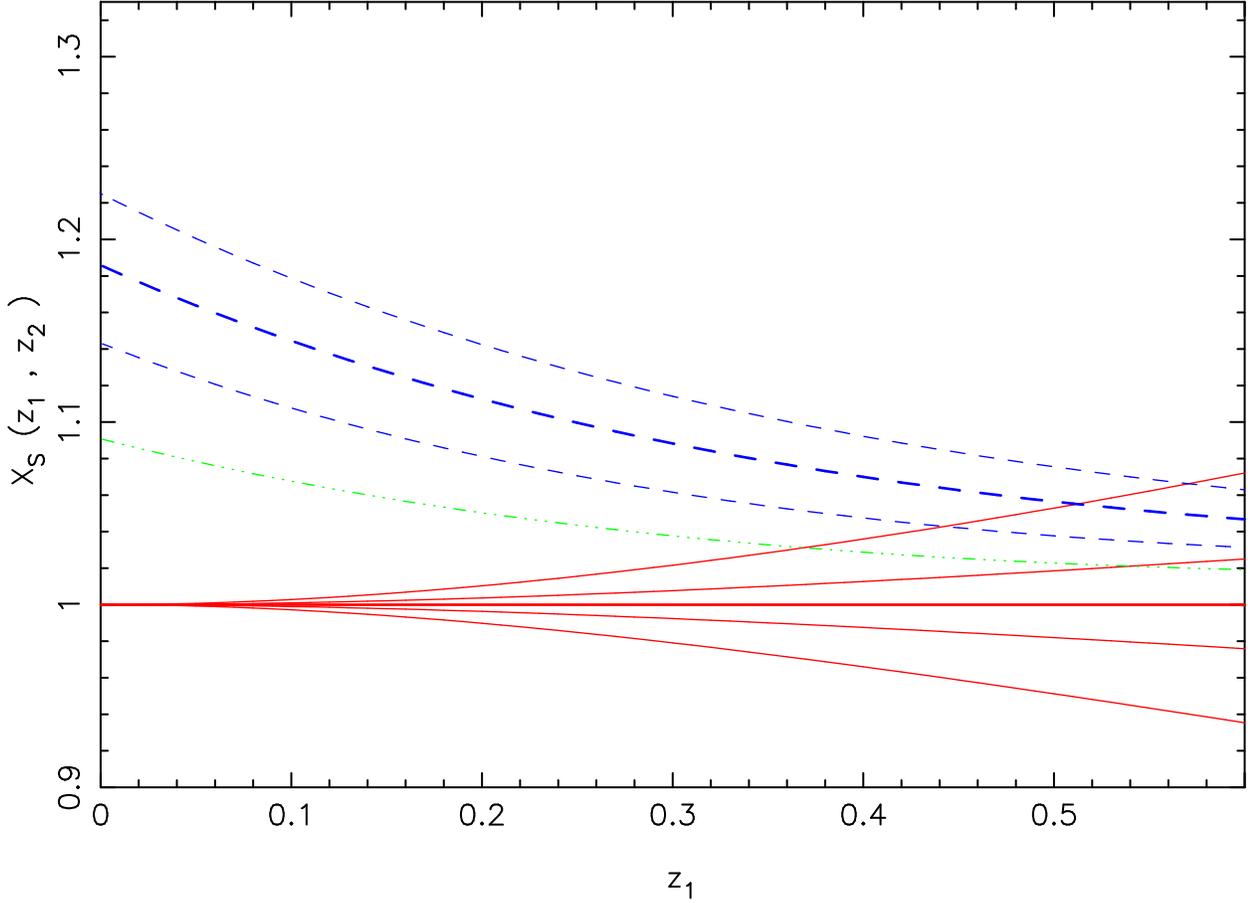} 
\caption{
 This figure shows both sides of inequality \ref{eq:xs} 
 as a function of redshift $z_1$. The solid lines show the
  right-hand side of \ref{eq:xs}, i.e. the upper limit
  on $X_S$ for non-accelerated models, assuming respective
 curvature radii $R_C = -0.6$, $-1.0$, $\infty$, $+1.0$, $+0.6$ in units
  of $c/H_0$  (bottom to top). 
The dashed lines show the predicted values of $X_S(z_1, z_2)$
 for varying $z_1$ at fixed $z_2 = z_{acc}$, 
 for the same four models as in Fig.~\ref{fig:hz}.  
 The three dashed lines show flat $\Lambda$CDM models 
  with $\Omega_m = 0.24$ (upper), 0.27 (thick), 0.31 (lower).  
The dot-dashed line shows wCDM with $\Omega_m = 0.32$,  $w = -0.85$.  
 }  
\label{fig:xs} 
\end{figure*}

In inequality (\ref{eq:xs}), the LHS $X_S$ is formed from a ratio of 
  two dimensionless BAO observables $d(z_1)$ and $y(z_2)$,
  while the RHS is close to 1 with a weak dependence on curvature:  
 the effect of curvature on the low-redshift $D_V(z_1)$ is
 folded into the factor containing $S_k$ on the RHS.     
 As noted above, this is exactly 1 for flat models 
 and is always $< 1$ for open models,
  so open models always tighten the speed trap.   
 For closed (positively curved) 
  models the $S_k$ factor is $> 1$, which weakens
 the trap slightly; however, at low redshift $z_1$ this is 
 a small effect as follows: 
{\newtwo from the discussion in \S~\ref{sec:dvlim}, 
 for closed models we found a conservative lower
 limit $R_C \ge 0.6 \, c/H_0$;  this leads to $x_1 \le \ln(1+z_1)/0.6 $, 
  thus for example the RHS is $\le 1.013$ for $z_1 = 0.2$. 
 The top solid curve 
  in Figure~\ref{fig:xs} shows the resulting upper limit 
  on the RHS of (\ref{eq:xs}) assuming the very 
  conservative limit $R_C \ge 0.6 \, c/H_0$,
  while the next-to-top
 solid curve shows the limit assuming $R_C \ge 1.0 \, c/H_0$. } 

 Thus, if actual observations reveal that $X_S \simgt 1.02$ with
  good significance,  the cosmic speed-trap ``flashes'': if so,  
 we can then rule out all homogeneous non-accelerated models
 regardless of the detailed expansion history or gravity model. 

 In the above Eq.~\ref{eq:xs},  the square bracket term in $X_S$  
 is given to first order by $(1 + \frac{2}{3}z_1)^{-1}$. 
  Higher order terms
  are small, and a quadratic approximation is not an improvement; 
 a slightly better approximation is $(1 + 0.65 \, z_1)^{-1}$
 which is accurate to $0.2 \%$ for $z_1 \le 0.3$. 
 Note that the RHS of (\ref{eq:xs}) has no dependence on $z_2$; 
 the curvature radius $R_C$ has no effect on the observable $y(z_2)$
  since $y$ is purely a line-of-sight measurement. 
 Therefore, we may choose to measure $y(z_2)$ anywhere, 
  but if the real universe is accelerating, the observed
  $X_S$ will be maximal when 
  $z_2$ is close to the past minimum of $y(z_2)$, 
 at $z_2 \approx \zacc$.

\subsection{Predictions for $\Lambda$CDM} 
\label{sec:pred} 

In Figure~\ref{fig:xs} we show 
  predictions for $X_S(z_1, z_2)$ as a function of $z_1$ 
  for three $\Lambda$CDM models (dashed) 
  and one wCDM model (dash-dot), from substituting Eq.~\ref{eq:dovery} 
 into \ref{eq:xs} and evaluating $H(z)$ and $D_V(z)$
  for the models.     For each of these plotted curves, 
   $z_2$ is set to $\zacc$ for that model.  
 The non-accelerating upper limit for
  $X_S$ (the RHS of Eq.~\ref{eq:xs}) is shown as
  solid lines for several assumed values of curvature radius $R_C$.

We see from Figure~\ref{fig:xs} that if the real universe has followed 
 an expansion history $H(z)$ similar to $\Lambda$CDM prediction, 
 inequality \ref{eq:xs} will be violated if $z_1$ is reasonably small
 and $z_2$ is near $z_{acc} \sim 0.75$.   
 Essentially, the accelerated expansion between $z_2$ and $z_1$ 
 causes the value of $H(z)/(1+z)$ to be larger at $z \le z_1$ than 
 in the past at $z_2$, as in Figure 1; this makes $D_V(z_1)$ 
  smaller and $d(z)$ larger, compared to any non-accelerating model
  with the same $H(z_2)$, 
 so $X_S$ violates the limit in Eq.~\ref{eq:xs}. 

 As noted above, to maximise the violation we should choose $z_2$ to 
  minimise the observed value of $y(z_2)$, i.e. the redshift $\zacc$ 
 where  $H(z)/(1+z)$ had its past minimum; for a $\Lambda$CDM model with
 $\Omega_m = 0.27$, the actual minimum is at 
 $\zacc \approx 0.75$, but the theoretical $y(z)$ is 
  within 2\% of its minimum value over a rather broad window 
  $0.5 \simlt z \le 1.1$: so for an observational application of
  the test, 
  $z_2$ may be whatever is most convenient observationally
 within this range, with only marginal weakening of the trap.  
 
 Turning to the variation of $X_S$ with $z_1$, 
  the predicted value of $X_S$ is maximal at $z_1 = 0$
 (with a value of 1.185 for our reference model C), and slowly 
  declines with $z_1$: thus lower $z_1$ is better both to
  maximise lever-arm in our speed-trap, and to minimise
 curvature uncertainty. 
  However, for practical observations $z_1$ cannot be too small 
 since we need sufficient cosmic volume to get a robust 
 detection of the acoustic feature in 
 the galaxy correlation function $\xi(r)$ or power spectrum $P(k)$; 
 therefore there is a tradeoff between $X_S$ which declines with $z_1$, 
 the curvature uncertainty also favours smaller $z_1$, but  
 the available cosmic volume for measuring $d(z_1)$ 
 grows with $z_1$.  Thus for an observational application
 of the speed-trap, there is an optimal
 window around $0.15 \simlt z_1 \simlt 0.35$. 

 Taking example values $z_1 = 0.1$, $0.2$, $0.3$,  the concordance model 
  predicts $X_S(z_1, \,0.75) = 1.145$, $1.113$, $1.088$ respectively.   
 We also note that the value of $X_S$ is fairly  
 sensitive to the value of $\Omega_m$: taking example cases from 
 Table~\ref{tab:models} 
 with $\Omega_m = 0.24, 0.27, 0.31$ to bracket the plausible
 range, we find that $X_S(0,\zacc) = 1.225$, 1.185, 1.143 respectively; 
 while $X_S(0.2, \zacc)$ is 1.142, 1.113, 1.081.  
 For each model, $X_S - 1$ approximately halves 
  from $z_1 = 0$ to $z_1 \approx 0.27$.  
 This is because the rate of acceleration grows with time after $\zacc$, 
  so $X_S$ has stronger than linear dependence on $q_0$. 
 
{\newtwo 
We note here that the prediction for $X_S$ is independent of $H_0$ if all 
 of $\Omega_m$, $\Omega_{DE}$, $\Omega_k$ and $w$ are held fixed. 
 However, since our example models
 are approximately CMB-matched, a correlation appears, because
  raising $\Omega_m$ and/or $w$ compared to the
 concordance model requires lowering $H_0$
 to remain consistent with the CMB;  while 
  raising $\Omega_m$ or $w$ also leads to weaker acceleration and thus 
  lowers $X_S$. Thus, $X_S$ at a fixed redshift 
  is positively correlated with $H_0$ in CMB-matched Friedmann models.   
} 

We also note that for accelerating models $X_S$ remains 
 a few percent greater than 1 for the case $z_1 = z_2$; this occurs because 
 $y(z_2)$ measures the instantaneous expansion rate at $z_2$, while 
  $d(z_1)$ depends on the average expansion rate at
   redshifts below $z_1$, which is larger. In principle we could 
  use this to test acceleration by measuring 
  $d$ and $y$ from a single survey at $z_1 = z_2$, 
 but in practice the curvature uncertainty probably disfavours this 
 (see Sec~\ref{sec:ap} for more discussion). 
 
\section{Discussion}
\label{sec:disc}  

In this section we discuss various aspects of the test above,
 including possible shifts in length $r_s$, useful approximations
 for $D_V(z)$, observational issues, 
 the relation to the Alcock-Paczynski ratio and 
  the effect of giant-void models. 

\subsection{Possible shifts in $r_s$} 
\label{sec:rshift} 

In applying the speed-trap, clearly assumptions (i) and (ii)
 above are very basic; if 
 future observations show the speed-trap is observationally violated,  
 we need to be confident that assumption (iii) on constancy of
  $r_s$ is valid to around $\sim 2\%$, in order to reject general 
 homogeneous non-accelerating models with high confidence. 

We now consider some details which may actually give rise to
 a significant shift in comoving $r_s$ between redshifts $z_1$ and $z_2$; 
the main such effects are galaxy bias, non-linear growth of structure, 
  redshift-space distortions \citep{kaiser87, ham92},  
 and possible effects due to the hump(s) sitting on a sloping
 background power spectrum etc. 

{\newtwo 
We first note that there is non-negligible evolution in $r_S$ 
 at high redshift between 
 $z_d \approx 1020$ and $z \sim 10$, as shown by Figure~1
 of \citealt{esw07}; the initial BAO bump 
 is only in the baryons and photons, 
 and the peak shifts slightly as the dark-matter
 and baryon perturbations align together at later times;  
  this implies the late-time BAO peak is not exactly at 
  the sound horizon length $r_S(z_d)$.  
 However, after $z \simlt 10$ the density perturbations in
  baryons and dark matter are very similar. In most real BAO 
 analyses, a matter power spectrum from CMBFast or similar 
  is used, together with a model for non-linear
 evolution and an arbitrary linear ``stretch factor'' $\alpha$, 
 to fit observations;  
  finally, the measurement is quoted as $d(z) = \alpha \, r_S(z_d) / D_V(z)$
  where $r_S$ and $D_V$ are both computed from the reference 
  theoretical model.
 This implies that small errors in the reference model should (on average) 
  be absorbed into an opposite shift in $\alpha$, so 
   the final estimate of $d(z)$ should be unbiased. 
  Any shift in the BAO length from $z \sim z_d$ to $z \sim 10$
   is included in the reference model; 
  therefore, $r_S(z_d)$ forms essentially 
 a convenient fiducial length for intercomparison between models, which is
  close to but not exactly the position of the 
 low-redshift BAO peak in the correlation function.
 For the present work, we are only interested in shifts of the BAO scale
  at $z < z_2 \sim 0.75$, so the above effect cancels.  
 } 

Galaxy bias, at least in standard versions, has little
 effect since the BAO scale is very much larger than any scale 
 of relevance for galaxy formation; thus bias may affect the
 overall amplitude of galaxy clustering but cannot significantly
 shift the scale $r_s$. 
 Likewise, non-linear growth of structure primarily
 moves galaxies around on $\sim 5 \hmpc$ scales; this significantly 
  blurs the bump in $\xi(r)$, and/or erases the higher harmonics
 in the power spectrum, but this is almost symmetrical
  between inward and outward shifts: the systematic shift 
in the BAO lengthscale is much smaller. 

For the standard model, these effects have been investigated
 from both theory by e.g. \citet{esw07} and \citet{shoji09},  
  and from large N-body simulations by 
 \citet{seo08} and \citet{seo10}; these papers agree 
 that systematic shifts are small, typically below the $0.6\%$ level at 
 $z = 0.3$ and less at higher redshift. They also find that
 reconstruction methods based on velocity-field reconstruction
  \citep{esss07}  
  can reduce the shift to $\sim \, 0.1\%$. 
  This will become important
 for the next generation of ambitious planned surveys such as ESA's Euclid
 (e.g. \citealt{sam11}) or NASA's WFIRST, which aim
 to achieve sub-percent precision on BAO observables in many 
  redshift bins, but are almost negligible
  with respect to the speed-trap test in this paper. 

 We caution that there is a slight level of circular 
  argument in the above, in that
 we are assuming standard cosmology to limit the shift in $r_s$, and
  then using this to reject non-standard non-accelerating models; 
 it remains possible that a model with non-standard gravity could
 produce a much larger shift in $r_s$ than the standard cosmology. 
 However, non-standard models producing a gross 
   $\sim 10\%$ shift in $r_s$ 
 since $z_2 \sim 0.75$ would almost certainly produce large levels of 
 redshift-space distortion, and give strong inconsistencies between the
 angular and radial measurements of $r_s$ at low redshift.  
 If both the redshift-space distortion pattern and the radial and angular
  measurements 
 of $r_s$ are measured to be consistent with standard $\Lambda$CDM, 
 this would strongly suggest that the true shifts in $r_s$ should not
 be much larger than the percent level effects predicted by the
  standard model.

\subsection{Approximations for $D_V$} 
\label{sec:dvapprox} 

As an aside, we also note that in nearly-flat CDM-like models, 
 an accurate approximation to $D_V(z)$ at moderate redshift 
 is given by Taylor-expanding $1/H(z)$ around $z/2$ (rather than zero), and
 substituting in the integral Eq.~\ref{eq:dr};
 this makes $z^2$ terms vanish, and leads to the approximation 
\begin{equation} 
  D_V(z) \approx {c z \over [H^2(z/2) \, H(z)]^{1/3} }  + O(z^3)  \ ;
\label{eq:dvapp1} 
\end{equation} 
 in practice the first term is surprisingly accurate for $\Lambda$CDM 
 models, with errors $< 0.1\%$ compared to the numerical
  result for $z < 0.5$. (See Appendix A for evaluation of the
 third-order term, and explanation why it is small). 

A simpler approximation is 
\begin{equation} 
  D_V(z) \approx {c z \over H(2z/3) } \ ;
\label{eq:dvapp2} 
\end{equation} 
 this is slightly less accurate than the previous approximation, 
  but still accurate to $< 0.4\%$ for $z < 0.5$, better than the
 mid-term precision on observables.  
 (For open zero-$\Lambda$ models
  these approximations are less good, with errors up to $2\%$). 

While it is straightforward to evaluate $D_V$ and $X_S$ 
 numerically for any given model,  
  the main value of this approximation is that it tells us
 that a measurement of $z\,d(z)$ at low redshift is quite close to  
  a measurement of $r_s H(2z/3) / c $ ; 
substituting this into (\ref{eq:xs}), along with 
 the approximation $(1 + \frac{2}{3}z_1)^{-1}$ for the
 square-bracket term,  gives simply  
\begin{equation} 
 X_S(z_1, z_2) \approx  { (1+z_2) \over H(z_2)}
  {  H\left(\frac{2}{3}z_1\right) \over 1 + \frac{2}{3}z_1 } 
   = { \dot{a}(\frac{2}{3}z_1) \over \dot{a}(z_2) }  \ ; 
\end{equation} 
and inequality \ref{eq:hineq} tells us this should be less than  
 1 for non-accelerated models.  
 Unlike our upper limit Eq.~(\ref{eq:xs}) this expression is
  not rigorous, but this gives a simple and fairly
  accurate approximation for
 what $X_S$ is measuring, i.e. it is closely related
 to the ratio of expansion rates $\dot{a}$  at $\frac{2}{3}z_1$ compared
 to $z_2$. 
  
\subsection{Observational advantages} 

One possible objection to this test is that it is comparing two
 related but  slightly different observables, i.e. a spherical-average scale
 at $z_1$ with a radial scale at $z_2$.  Why have we done this, 
 rather than comparing two measures of $d(z)$ or two measures of $y(z)$ 
 at two different redshifts ? 

 It is well known that comparing $y(z)$ at two different 
 redshifts provides
 another direct test of acceleration. 
 The main difficulty is observational, 
  since for our baseline model,
  $y(z_1)$ only grows to $1.1 \, y(z_{acc})$ at rather low redshift
  $z_1 \sim 0.16$. Furthermore, for a given survey, a
  radial-only measurement of $r_s$ has a statistical error
  roughly $\sqrt{3}$ worse than a spherical average measure. 
  Even if we had a $3 \pi$ steradian redshift survey at 
 $z_1 \approx 0.16$, we may not do much better 
 than 3\% statistical error on $y$, 
  a $3\sigma$ violation,     and we would like to get above 
  $5\sigma$ for a decisive result.   
 Using $d(z_1)$ instead of $y$ gives 
  two substantial advantages: firstly $d(z_1)$
 effectively measures $H$ at $\sim 2 z_1/3$, giving
 more lever-arm on the low-redshift acceleration;  
 so a measurement of $d(z_1 = 0.24)$ is similar in content  
  to a measurement of $y(z_1 = 0.16)$. Secondly there is
 the obvious gain that $d$ uses 3 spatial directions instead of 1. 
 Thus for a fixed thickness of survey shell, the former measure 
 has around 9/4 times more available volume and 3 independent axes,
 so the cosmic variance limit should improve 
 by a factor $\sim \sqrt{27/4} \approx 2.6$, which is a very important 
 practical advantage.   

In contrast, comparing $d(z)$ at two different redshifts suffers from
 potential major uncertainty in cosmic curvature at the high redshift $z_2$.
 At $z_2 \approx 0.75$ there is ample available volume for a 
 precision measurement of $y$, and ambitious future probes 
 such as Euclid (\citealt{sam11}) plan 
 to push to statistical errors $\simlt 0.75\%$ on 
  both $y$ and $d$, in each of many bins of width 0.1 in redshift. 
  Thus, at $z_2$ the cosmic variance is minimal for a wide-area
 survey, so the radial measure
 is preferable because it is independent of the curvature nuisance parameter. 
  Also, $d(z_2)$ depends on the full history of $H(z)$ back 
  to $z_2$, which complicates the issue of deriving an inequality. 
   
 In our proposed comparison, we have constructed a 
 ratio $X_S$ using $d(z_1)$ at low 
 redshift and $y(z_2)$ at the higher redshift, to circumvent both 
 of these problems: 
 the potential cosmic-variance limits are probably around $1\%$ 
  on $d(0.24)$ and significantly less on $y(0.75)$, so this test can 
 (given ample data) deliver a standalone rejection of homogeneous
  non-accelerating models at $\sim 7\sigma$ significance level. 
 This can be further improved by using several independent redshift
  bins, e.g. $z_1 = 0.15, 0.25$ and $z_2 = 0.65, 0.75$.

\subsection{Future Observations} 

As noted above, there already exist measurements of the 
 numerator on the left of Eq.~\ref{eq:ydlim} from \citet{perc10};
 they quote values of $d(0.2) = 0.1905$ and $d(0.35) = 0.1097$,  
  with approximately 3.3\% error on each. 
 For the numerator $z \,d(z)$ in inequality (\ref{eq:xs}) 
  these give $0.2 \,d(0.2) = 0.0381$ and $0.35 \,d(0.35) = 0.0384$. 

As yet there is no available measurement of radial BAOs at $z > 0.5$ 
 with which to actually calibrate our speed-trap, but 
 these are expected soon 
\footnote{ Soon after the submission of
 the first version of this paper, three new measurements of the BAO 
 feature appeared: one from 6dFGS at $z \sim 0.1$ 
  in \cite{beutler11}, one from WiggleZ at $z = 0.6$ in 
  \cite{blake11}, and one from SDSS photo-z's at $z = 0.55$
 in \cite{carnero11}. All of these show good consistency 
 with the concordance model, but do not yet measure the radial 
 component as required here.} 
 from the recently completed
 AAT WiggleZ survey \citep{blake10}, 
  and in a few years from the ongoing BOSS survey \citep{white11}. 
  It is currently unclear whether the 
 final WiggleZ survey covers enough volume to separately measure the radial
 component as required here, but BOSS should very likely achieve this; 
 the upper redshift limit of BOSS is $\approx 0.65$, so this is close
  enough to $\zacc $ to be useful. 
 
 For $\Lambda$CDM, the predicted value of $y(z_2)$ near its minimum
  is approximately 0.0302 for $\Omega_m = 0.27$ and 
 $H_0 = 70 \hunit$. 
 For reasonable variations of parameters, 
 we now show that if we assume a flat universe
  then $y(z_2)$ is well constrained by CMB observations: 
 it is well known that for flat models with varying $\Omega_m, h, w$ 
  there is a tight correlation
 between the age of the Universe, $t_0$, and the CMB acoustic scale $\ell_A$ 
 \citep{kcs01}; and 
 it turns out that there is also a tight correlation between these
 and the value of $H$ at intermediate redshift, with a pivot point
  occurring at $z \approx 0.8$ (see Figure~\ref{fig:hz}).  
 This is partly a coincidence, because for
  moderate parameter variations around the concordance model, $t_0$
  scales $\propto \Omega_m^{-0.3} h^{-1}$, 
 while $\ell_A$ scales as $\Omega_m^{-0.15} h^{-0.5}$.   
  For the value of $H(z)$, 
  we note that as $z \rightarrow 0$ this scales as $h$ independent
  of $\Omega_m, w$, while
 at $z \simgt 3$ where dark energy is negligible, $H(z)$ scales 
  $\propto \Omega_m^{0.5} h^1$. Therefore, there exists a pivot-point 
  at intermediate
  redshift where $H(z)$ scales as $\Omega_m^{0.3} h$ (i.e. inversely
  to $t_0$), and this pivot redshift turns out to 
   be $z \approx 0.85$ for $\Lambda$ models.  For $w > -1$
 the pivot redshift is somewhat lower, but  
  for near-flat Friedmann models the value $H(z = 0.75)$ is 
 better constrained by WMAP data than the local $H_0$;  
 and fixing $t_0 \approx 13.75 \, {\rm Gyr}$ 
 constrains $H(z = 0.75)/(1.75) = 59.2 \hunit$ within $\approx 0.8\%$,
 which in turn leads to a tight prediction for $y(0.75)$.  

 (As an aside, there is a corollary 
  that if some future method could give a direct measurement of 
  $H(z = 0.75)$ independently of $r_S$, this would produce another strong
  consistency test of standard $\Lambda$CDM. 
 This may be possible
  in principle using methods such as differential-age measurements of
  early-type galaxies,  or lensing measurements with source
 and lens close in redshift, but this will require a major advance
  in precision over current data). 

 Assuming some future $y$ measurement turns out at the concordance
 value $y(z_2) \approx 0.0302$,  
  we would then obtain measurements 
  $X_S \approx 1.12$ and 1.05 at $z_1 = 0.2$, 0.35 respectively.    
   The error on $y(z_2)$ must be added in quadrature to the current
  3.3\% error on $d(z_1)$,   but if the former is around 2\% then 
  we can anticipate a fairly clear violation from the $z = 0.2$ value,
  and a somewhat less significant violation at $z = 0.35$.  

 The prospects are good for improving on the current results:
 the projections for the BOSS survey \citep{white11} are 
 for $1\%$ precision on $d(z = 0.35)$, 
 and precision of $1.7\%$ on $y(0.6)$. 
 Adding the above errors in quadrature leads to around 2\% 
  precision on $X_S$, with a predicted value $\approx 1.077$, thus
  nearly a $4\sigma$ proof of acceleration. BOSS may also do better
  using the larger value of $X_S$ at $z_1 \sim 0.2$, but projected 
 precision on $d(0.2)$ is not quoted separately.   

Next-generation surveys in the planning stage such as
 BigBOSS, Euclid or WFIRST should substantially improve on the
 higher-redshift measurement, reaching sub-percent precision on
 $y(z_2)$.  The low-redshift $d(z_1)$ measurement is ultimately
 limited by cosmic variance, but extending the BOSS survey
 to the Southern hemisphere can give a straightforward improvement
 by a factor of $\sqrt{2}$, or probably more if denser sampling of
 galaxies is used.  Further improvements are possible
 in principle using HI or near-infrared selected 
  surveys which can cover $> 80\%$ of the 
  whole sky, compared to $\sim 50\%$ for visible-selected surveys. 

\subsection{Comparison with the Alcock-Paczynski test} 
\label{sec:ap} 

{\newtwo 
 We note here that our ratio $X_S$ may be considered as
 a generalised version of the
 classic test of \cite{alc-pac}, hereafter AP:  
 the AP ratio was defined to be $R_{AP} \equiv \Delta z / z \Delta \theta$, 
 which in our notation becomes 
\begin{equation}
 R_{AP}(z) \equiv { (1+z) D_A(z) H(z) \over cz } \ . 
\end{equation}  
If we choose $z_1 = z_2$ in Eq.~\ref{eq:dovery} above
 and substitute Eq.~\ref{eq:dv} for $D_V$, we then obtain 
\begin{equation} 
  { z_1 d(z_1) \over y(z_1) }  = (1+z_1) R_{AP}(z_1)^{-2/3} 
\end{equation} 
 thus $X_S(z_1, z_1)$ contains the same information as $R_{AP}(z_1)$ 
  combined with a function of $z_1$;  substituting the above into 
 Eq.~\ref{eq:xs} 
 gives a lower limit on $R_{AP}$ for non-accelerating models,
  which is 
\begin{equation}
\label{eq:aplim} 
 R_{AP}(z_1) \ge { (1+z_1) \ln (1+z_1) \over z_1 } {S_k(x_1) \over x_1} \ .
\end{equation}  

It is well known that if we assume the Friedmann equations, 
 the AP test at high redshift provides a strong test
 for $\Lambda$ or dark energy: however, 
  if we drop the Friedmann connection between curvature
 and matter content,  then at $z \simgt 0.5$ the AP test becomes mostly
 degenerate between acceleration and curvature.  
At lower $z < 0.4$,  we may use the approximation $(1+z) D_A(z) 
  \approx cz /H(z/2)$
  from above, which leads to 
 $R_{AP}(z_1) \approx H(z_1) / H(\frac{z_1}{2})$.
  This does have more
 sensitivity to acceleration than curvature, but is not ideal
 for the following reason:
  at small $z_1$ the AP ratio suffers from a short redshift lever-arm,
 while at $z_1 \simgt 0.4$ the ratio mainly probes the regime of sluggish
  acceleration at $z > 0.2$.  The AP ratio at $z_1 \approx 0.4$
 may provide a useful test, but will probably require sub-percent
  level precision on both observables to get a decisive result. 

Compared to the AP test, 
 the use of two widely-spaced redshifts in $X_S$ requires the
 added assumption that $r_S$ has minimal evolution 
  between $z_2$ and $z_1$,
  but enables a much longer effective time lever-arm,  
  giving a larger acceleration signal 
  while keeping the curvature sensitivity very small. 
} 

\subsection{Inhomogeneous Void Models} 
\label{sec:void} 

{ \newtwo 
Recently there has been some interest in models which produce
 apparent acceleration without dark energy, by placing us near the centre of
 a giant underdense spherical void, with a Lemaitre-Tolman-Bondi metric;
 examples are in \cite{tomita09} and references therein.  
 These models have several problems such as severe fine-tuning
 of our location very close to the void centre, and probable
 inconsistency with limits on the kinetic Sunyaev-Zeldovich
  effect \citep{zhang11};
  however it is interesting to note how $X_S$ behaves in such models.  
 A recent confrontation of giant-void models with BAO observables
  has been done by \cite{moss11}: they find that  
 void models with profiles adjusted to match SNe and CMB observations 
 have a $\Delta z_{/\!/}$ which is $\simgt 30\%$ smaller at $z \sim 0.5 - 0.7$ 
  compared to $\Lambda$CDM. 
 Those specific cases would have $X_S(0.2, 0.75) \simgt 1.4$,  
  which is substantially larger than any reasonable dark-energy model; 
 thus,  \cite{moss11} show that giant-void models
 matched to angular distances and the CMB 
 appear to suffer from severe ``overkill'' in 
  radial BAO measurements.     

 The parameter space of possible void models is very large, so other
  void models may look more similar to $\Lambda$CDM, 
   but we note that the test of \cite{cbl08} 
  can be used to test for 
 homogeneity {\em without} assuming GR. They show that if
  we have both angular and radial BAO measurements 
  spanning a range of redshift, there is a consistency relation
  which must be satisfied by homogeneous models but is usually
 violated by giant-void models. Thus, assumption (i) above
  becomes observationally testable using future BAO observations,
  though this probably requires observations spanning more redshifts 
  than the $X_S$ test here.  
 } 

\section{Conclusions}
\label{sec:conc} 

We have proposed a new and simple smoking-gun test for cosmic acceleration 
 using only a comparison of the baryon acoustic oscillation feature
 at two distinct redshifts $\sim 0.2$ and $\sim 0.75$. The main result of
 our paper is inequality~(\ref{eq:xs}) relating
 the two dimensionless BAO observables, which must be satisfied
 for any homogeneous non-accelerating model, but will be
 observationally violated by $\approx 10\%$ in models with an
  expansion history close to standard $\Lambda$CDM.  

Clearly, our proposed measurement has advantages and disadvantages:  
 the main advantages are extreme simplicity and model-independence,
 i.e. if the inequality (\ref{eq:xs}) is violated, we
 can rule out essentially all homogeneous non-accelerating models 
 in one shot, without assuming any particular gravity theory or
  parametric form of $H(z)$, {\newtwo and independent of supernova
 and CMB observations. }  

The main drawback of our test is that it is essentially one-sided: 
 if inequality \ref{eq:xs} is observationally violated, 
 we have proved (given some basic assumptions) 
 that acceleration has occurred during 
  $0 \le z \le z_2$ and have a rough quantification of the amount, but 
 no more details about the underlying cause or the details of the 
 expansion history. 

If we assume GR and the Friedmann equations hold, and
 that $r_s$ has the value which is accurately 
 predicted from CMB analysis, then we have much more statistical power: 
  future measurements of BAOs in many redshift bins may be used to
  reconstruct the detailed form of the expansion history $H(z)$;
 this can be integrated to give predictions of $D_R(z)$, and comparison
 with the measured transverse BAO scale giving 
  $D_A(z)$ can constrain spatial curvature independent of the CMB;  
 while comparison of $D_A$ with $D_L(z)$ from SNe can check the 
  distance-duality or Tolman relation $(D_L/D_A)^2 = (1+z)^4$.  
  All of this can give much more 
   powerful cross-checks and parameter estimates
  than our simplified one-sided test. 

However, our proposed cosmic speed-trap seems to provide a 
 valuable addition to the set of cosmological measurements, due to its 
 bare minimum of assumptions.  This provides a
 strong motivation for future
 improved BAO measurements specifically near 
 redshifts $\sim 0.25$ and 0.75; 
 this should  preferably include a low-redshift survey 
 comparable or superior to BOSS in the Southern hemisphere 
 to minimise the cosmic variance in the local measurement.

\section*{Acknowledgments}

 I thank Steve Rawlings for a perceptive question which 
 spurred this investigation, and I thank John Peacock 
 and Will Percival for helpful discussions which improved the paper.  
 I also thank Jim Rich for discussion on curvature limits, and 
  I thank the referee for several helpful clarifications. 


\appendix

\section[]{The approximation for $D_V$}

We here add a note which explains the surprisingly good
 accuracy of approximation \ref{eq:dvapp1} for $D_V(z)$ at fairly low
 redshift $z < 0.4$.   
As noted,  in the integral Eq.~\ref{eq:dr} for $D_R$, 
 it is helpful to Taylor-expand the function $1/H(z)$ 
 around the mid-point of the integral at $z_1/2$, then 
 integrate: this naturally
 makes terms with odd-integer derivatives of $1/H$ 
  integrate to zero, and leads to 
\begin{equation}
 D_R(z_1) \approx  c \left[ {z_1 \over H(\frac{z_1}{2}) }  + {z_1^3 \over 24 } 
    \left( { 1 \over H}\right)'' (\frac{z_1}{2}) 
  + O(z_1^5)  \right]  
\end{equation} 
 where prime denotes $d/dz$.  We now need the second derivative $(1/H)''$
 evaluated at $\frac{z_1}{2}$. 
 Defining the usual deceleration parameter $q$ and the jerk parameter $j$ 
 (e.g. \citealt{alam03}) as  
\begin{equation} 
 q \equiv -{d^2 a/dt^2  \over a H^2} \ , 
  \qquad j \equiv { d^3a / dt^3 \over a H^3} \ , 
\end{equation} 
  we can rearrange these in terms of $d/dz$ to get 
\begin{equation} 
  {dH \over dz} = {H \over 1 + z} (1 + q) \ , \qquad 
  {d^2 H \over dz^2} = {H \over (1+z)^2} ( j - q^2 ) \ .  
\end{equation} 
Using these we obtain  
\begin{equation} 
\label{eq:ddhinv} 
  {d^2 \over dz^2}\left( \frac{1}{H} \right) = 
   { -j + 2 + 4q + 3q^2  \over (1+z)^2 \,H } 
 \ . 
\end{equation} 
 
For the case of flat $\Lambda$CDM models,
  $j = +1$ independent of parameters (assuming
 radiation density is negligible) \citep{rapetti07}, 
 thus the numerator in Eq.~\ref{eq:ddhinv} 
 has zeros at $q = -1/3$ and $q = -1$. 
 For $\Omega_m$ near the concordance
 model, $q$ passed through $-1/3$ in the fairly recent past 
  at $z \sim 0.3$,  
 so the numerator is significantly smaller than unity 
 at low redshift. This explains qualitatively the very 
 good accuracy of approximation \ref{eq:dvapp1} near the
 concordance model, even up to significant redshifts $z \approx 0.5$.   

%
%


\bsp

\label{lastpage}

\end{document}